\documentclass[9pt,twocolumn,twoside]{ArXiVjnl}

\journal{josaa} 
\usepackage{color}

\setboolean{shortarticle}{false}
\DeclareRobustCommand{\rchi}{{\mathpalette\irchi\relax}}
\newcommand{\irchi}[2]{\raisebox{\depth}{$#1\chi$}}
\title{Lossless reshaping of structured light}
\usepackage{multicol,blindtext,graphics,graphicx,float,multirow}
\author[1]{Stirling Scholes}
\author[1]{Valeria Rodr{\'i}guez-Fajardo}
\author[1,*]{Andrew Forbes}
\affil[1]{University of the Witwatersrand, Structured Light Group, Physics, 1 Jan Smuts Avenue, Johannesburg, South Africa, 2000}

\affil[*]{Corresponding author: andrew.forbes@wits.ac.za}

\begin{abstract}
Structured light concerns the control of light in its spatial degrees of freedom (amplitude, phase and polarization), and has proven instrumental in many applications.  The creation of structured light usually involves the conversion of a Gaussian mode to a desired structure in a single step, while the detection is often the reverse process, both fundamentally lossy or imperfect.  Here we show how to ideally reshape structured light in a lossless manner in a simple two-step process.  We outline the core theoretical arguments, and demonstrate reshaping of arbitrary structured light patterns, in the process highlighting when the technique is applicable and when not, and how best to implement it.  This work will be a useful addition to the structured light toolkit, and particularly relevant to those wishing to use the spatial modes of light as a basis in classical and quantum communication.
\end{abstract}

\setboolean{displaycopyright}{false}
\begin{document}
\maketitle

\section{Introduction} \label{sec:Introduction}
Structured light refers to the tailoring or shaping of light in all its degrees of freedom \cite{roadmap}: time/frequency for temporal structuring of light, usually in the context of ultra-fast lasers, but more commonly spatial structuring of light in polarization, phase and amplitude, even with quantum states \cite{forbes2019quantum}, and made easy of late by the use of rewritable spatial light modulators for shaping light \cite{Forbes2016,SPIEbook,lazarev2019beyond}.  In this article we will consider the common case of spatial structuring of \textit{scalar} light, but emphasize that what we demonstrate here can be used for full control when used in combination with temporal \cite{weiner2011ultrafast,weiner2000femtosecond} and vectorial \cite{Chen2018,zhan2009cylindrical,rosales2018review} shaping tools.  Although it is possible to engineer a desired form of structured light directly at the source by custom lasers \cite{forbes2019structured}, most laboratory experiments require the conversion, or reshaping of one form of light into the desired structure.  Often the problem is the conversion of a Gaussian beam into some other form.  Many specific solutions exist for this, for example, conversion of Gaussian beams to flat-top beams with refractive \cite{hoffnagle2000design}, diffractive \cite{turunen1998diffractive} and holographic approaches \cite{ma2010generation}, the creation of vortex beams by spiral phase plates \cite{Beijersbergen1994}, geometric phase \cite{rubano2019q,brasselet2009optical} and spatial light modulators \cite{white}. 
It is not fully appreciated that when this is done in a single step, as with many of these examples, the transformation is either imperfect or lossy.

\begin{figure}[ht]
	\centering
	\includegraphics[width=\linewidth]{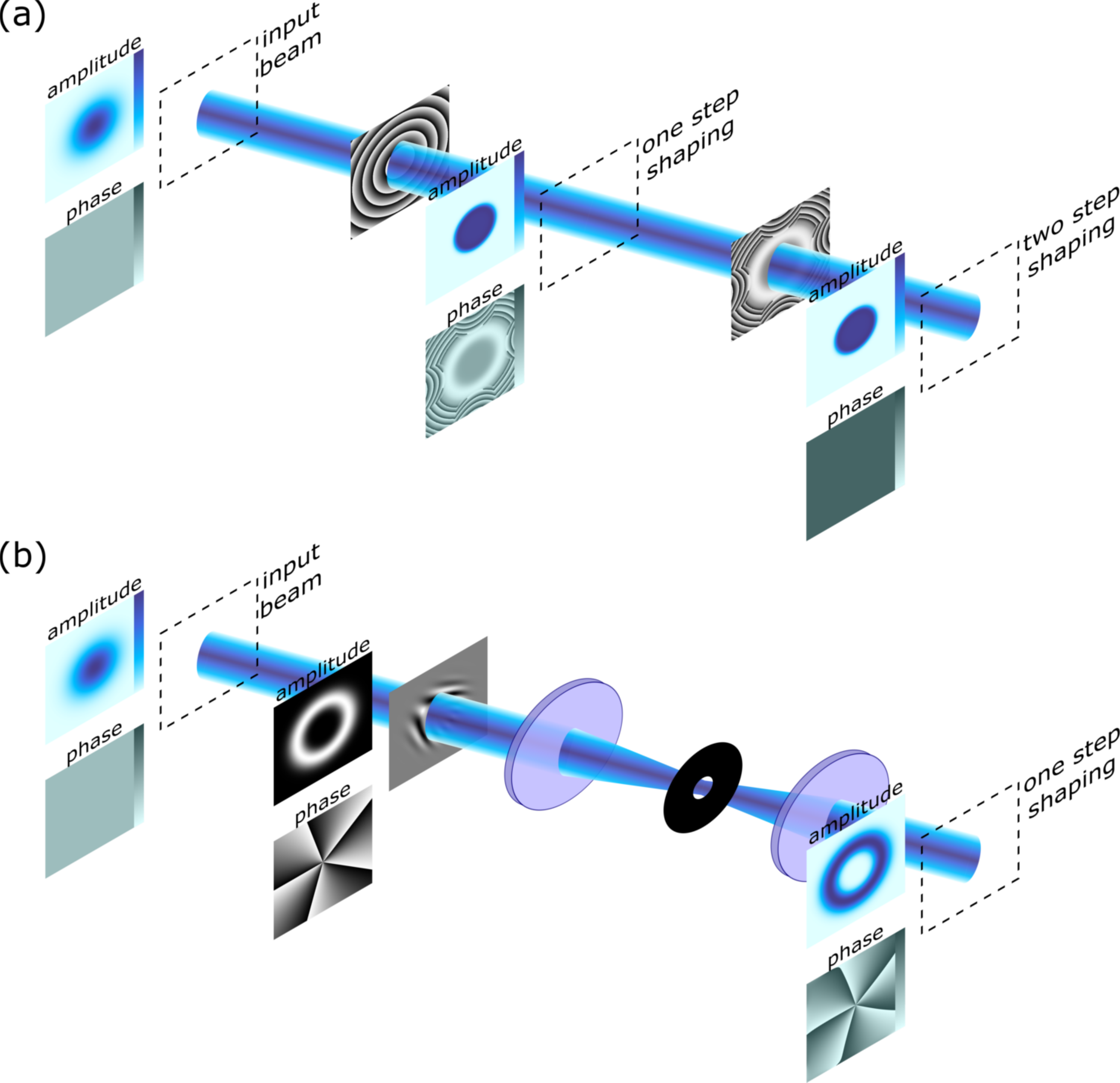}
	\caption{The different approaches to creating structured light. Panel (a) shows how lossless single step shaping can be achieved using a phase-only element, but requires the phase at the target plane to be left as a free DoF, requiring a second element for correction. Panel (b) shows how a single step in which the amplitude of the beam is also controlled can produce ideal structured light, but now the process is lossy.
	}
	\label{fig:IntroFigure}
\end{figure}

To illustrate this, let us consider perhaps the simplest beam shaping problem: reshaping of a Gaussian beam in size while maintaining the Gaussian amplitude profile and input phase.  A single step approach, with a single lens, will correctly resize the beam in a lossless manner, but alters the phase so that the shaping is imperfect: we must still correct for the phase degree of freedom (DoF).  The answer of course is to use two lenses, a two-step approach.  Such a telescope, appropriately designed, will correctly shape both DoFs in a lossless manner (in principle).  This principle is evident in all reshaping of structured light.  To return to our earlier examples (see Fig.~\ref{fig:IntroFigure}), it is possible to create a flat-top beam losslessly in a single step if the phase is left as a free DoF \cite{forbes2014laser}, while single step shaping of Gaussian beams into vortex beams that carry orbital angular momentum (OAM) \cite{padgett2017orbital,shen2019optical} through azimuthal phase transformation results in the correct ''vortex'' phase but leaves the amplitude as a free DoF: many radial modes are excited with low power content in the desired ring of light \cite{karimi2007hypergeometric,sephton2016revealing}.  This can be overcome and perfect reshaping achieved in one step, but then the process is lossy, requiring amplitude control, as has been done for the creation of radial mode controlled vortex beams \cite{rafayelyan2017laguerre}.  

Thus, in general, a two (or more) step approach is needed for complete and lossless reshaping of structured light.  By applying tools from lossless beam shaping theory \cite{bryngdahl1974geometrical,rhodes1980refractive,dickey2018laser}, solutions have been developed for specific cases, notably the special case of Gaussian to flat-top conversion \cite{kreuzer1969coherent,kreuzer1965Laser,frieden1965lossless,dickey1996gaussian} implemented by a variety of means \cite{shafer1982gaussian,han1983reshaping,kawamura1983simple,jahan1989refracting,dickey2018beam}.  Likewise specific solutions have been found for the reverse problem, the \textit{detection} of structured light, for OAM modes \cite{Berkhout2010}, Bessel modes \cite{Dudley2013,trichili2014detection}, Hermite-Gaussian modes \cite{zhou2018hermite} and Laguerre-Gaussian modes \cite{fontaine2019laguerre} with lossless approaches always making use of two or more steps rather than a lossy single step \cite{bouchard2018measuring}.  In this paper we revisit the fundamental principles of lossless beam shaping and outline a generic approach for arbitrary reshaping of structured light beyond the special cases already visited.  We provide a general recipe to follow and highlight under what conditions it is applicable.  We show the first general reshaping solutions beyond the simple Gaussian to flat-top converters and implement the solutions experimentally as digital holograms on spatial light modulators, elucidating the critical design and implementation steps to make this work.  In doing so we offer a holistic set of tools for designing lossless reshaping systems based on phase-only elements with potential applications including the creation and detection of specific modes of light for optical communication, and lossless control of light in metrology, laser materials processing, optical trapping and tweezing and novel intra-cavity beam shaping. 

\section{Theoretical outline}
As highlighted in the introduction, the topic of lossless laser beam shaping is a venerable one dating back to the 1960s.  Many ideas have been formulated since, and summarised in good textbooks on the subject \cite{dickey2018laser}.  Here we apply the concept of redistributing the energy of an input beam with profile $|I(\rchi)|^2$ in a precise way to form an output beam with a desired intensity profile $|Q(\rchi)|^2$ at the focal plane of a lens, i.e., in the far field. We outline this approach and in the process bring the recipe together in a manner that makes it easy to implement.  

The first element performs the energy redistribution and the second element corrects for the phase.  This is achieved by imparting the incident beam with a specific phase $\phi$, as determined by the desired output profile. To derive the necessary $\phi$ three steps are used. The first is a calculation of a scaling constant $A$. This constant is the ratio of the energy of the input beam to the output beam. Next is the formulation of an intermediary mathematical function $\alpha(\rchi)$, where $\rchi$ are the components of some coordinate system. This is a spatial mapping function which moves energy from a specific spatial region of the input beam to a specific spatial region of the output beam in a way consistent with the conservation of energy from the scaling constant $A$. It is the calculation of this mapping function which lies at the heart of the method, since once $\alpha$ is known it is possible to perform the final step, which is to derive the phase delay $\phi$ which must be introduced to a beam to achieve the desired output profile. We wish to make clear that the theory of this is known \cite{dickey2018laser}, and has been applied to the specific case of Gaussian to flat-top converters \cite{dickey1996gaussian,romero1996lossless}, but here we provide a general recipe in the context of structured light, indicating how and when it works (and when it doesn't), and make clear how the design steps impact on practical implementation of arbitrary reshaping of structured light, which we implement for the first time. 

\subsection{Derivation of equations}
Consider two planes in space, the first at coordinate $(x,y,z=0)$ and the second at a distance $(x,y,z=2f)$ where $f$ is the focal length of a lens. The coordinate system for both the global position and the plane at $z=0$ is $(x,y,z)$, while the coordinate system exclusively on the plane at $z=2f$ is $(X,Y,Z)$. The function $\alpha$ maps regions of the input plane to regions of the output plane meaning $\alpha(x,y)\rightarrow (X,Y)$. A beam of profile $|I(\rchi)|^2$ is incident upon the phase element $\phi$ positioned at $(x,y,z=0)$. At the second plane corresponding to the focal plane of the lens a beam of profile $|Q(\rchi)|^2$ is desired. To calculate the expression for the phase element $\phi$ the energy scaling constant $A$ must first be calculated. The constant $A$ is given by
\begin{equation}
\begin{aligned}
A = \frac{\int_{-\infty}^{\infty}|I(\rchi)|^2d\rchi}{\int_{-\infty}^{\infty}|Q(\rchi)|^2d\rchi}.
\label{eqn: Aformula}
\end{aligned}
\end{equation}
The next step is to derive a differential expression for $\alpha$. This expression will describe how the energy of the input beam at the plane of the phase element should be mapped to the focal plane of the lens to produce the desired output profile. Consider a field affected by the phase element $\phi$ passing through a lens, the field at the focal plane of the lens in 1-dimension is given by
\begin{equation}
\begin{aligned}
P(X) \propto \int_{-\infty}^{\infty}I(x)e^{i\phi(x)}e^{-i\frac{2\pi}{\lambda f}Xx}dx.
\label{eqn: lensPlane}
\end{aligned}
\end{equation}
Recall that $\alpha(x)\rightarrow X$.\\
\begin{equation}
\begin{aligned}
P(\alpha(x)) &\propto \int_{-\infty}^{\infty}I(x)e^{i\phi(x)}e^{-i\frac{2\pi}{\lambda f}\alpha(x)x}dx.\\
\end{aligned}
\end{equation}
Let 
\begin{equation}
\begin{aligned}
\phi(x) &= \frac{2\pi}{\lambda f}\Phi(x)\\
\beta &= \frac{2\pi}{\lambda f}
\end{aligned}
\end{equation}
\begin{equation}
\begin{aligned}
P(\alpha(x)) &\propto \int_{-\infty}^{\infty}I(x)e^{i\beta[\Phi(x)-\alpha(x)x]}dx.\\
\label{eqn: stationary phase integral}
\end{aligned}
\end{equation}
Equation~\ref{eqn: stationary phase integral} can be evaluated by the method of stationary phase as
\begin{equation}
\begin{aligned}
P(\alpha(x)) &\propto I(x)e^{i\beta[\Phi(x) - \alpha(x)x]}e^{i\frac{\pi}{4}}\frac{\sqrt{2\pi}}{\sqrt{\beta\Phi(x)^{''}}}\\
\label{eqn: stationary phase}
\end{aligned}
\end{equation}
with
\begin{equation}
\begin{aligned}
\frac{d}{dx}\left[\Phi(x)-\alpha(x)x\right] = 0.
\end{aligned}
\end{equation}
If we have devised $\phi$ correctly then up to a scaling constant we can state that 
\begin{equation}
\begin{aligned}
|P|^2 \propto A|Q(X)|^2.
\end{aligned}
\end{equation}
Thus
\begin{equation}
\begin{aligned}
|I(x)|^2\frac{2\pi}{\phi^{''}} &= A|Q(X)|^2\\
\frac{|I(x)|^2}{A|Q(\alpha(x))|^2} &=\phi^{''}.\\
\end{aligned}
\end{equation}
Note ~\cite{dickey2018laser}
\begin{equation}
\begin{aligned}
\frac{d}{d\alpha}\left[\frac{d}{dx}\left[\Phi(x)-\alpha(x)x\right]\right] &= \frac{d^2\Phi}{dx^2}\frac{dx}{d\alpha}-1\\
\frac{d^2\Phi}{dx^2}&=\frac{d\alpha}{dx}.
\end{aligned}
\end{equation}
This gives $\alpha(x)$ in 1-dimensional Cartesian coordinates as
\begin{equation}
\begin{aligned}
\frac{d\alpha(x)}{dx}&=\frac{1}{A}\frac{|I(x)|^2}{|Q(\alpha(x))|^2}.
\label{eqn: Cartesian Alpha}
\end{aligned}
\end{equation}
This equation describes how energy can be mapped from the input profile to the output profile in a manner consistent with the conservation of energy described by $A$.
The phase $\phi(x)$ can then be easily calculated from the expression for $\alpha$. The phase is
\begin{equation}
\begin{aligned}
\frac{d\phi(x)}{dx}&=\beta\alpha(x)\\
\phi(x)&=\beta\int_{0}^{x}\alpha(s)ds.
\label{eqn: dickey phase eqn}	
\end{aligned}
\end{equation}
The parameter $\beta$ provides an indication of how well the generated beam will conform to the ideal output profile. The greater the value of $\beta$, the higher the fidelity of the beam at the focal plane with regards to the ideal output profile.\\
\\
Solving the system of Eqns.~\ref{eqn: lensPlane}-\ref{eqn: Cartesian Alpha} in polar coordinates gives\\
\\
\begin{equation}
\begin{aligned}
\frac{d\alpha(r)}{dr}&=\frac{1}{A}\frac{r|I(r)|^2}{\alpha(r)|Q(\alpha(r))|^2}.
\label{eqn: Radial Alpha}
\end{aligned}
\end{equation}

\begin{figure}[H]
	\centering
	\includegraphics[width=.5\textwidth]{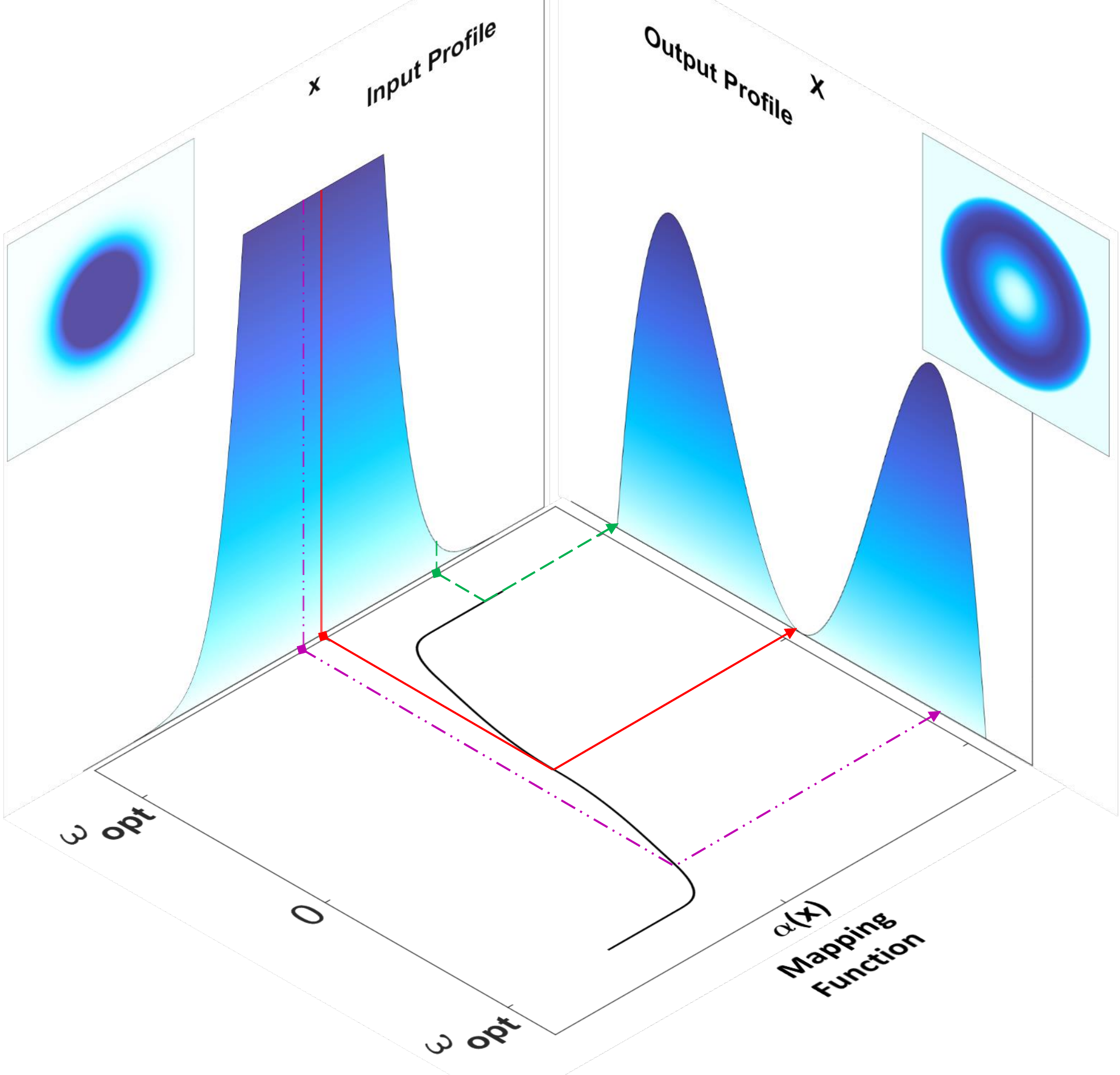}
	\caption{The mapping function $\alpha$ for a truncated Gaussian input to an annular polynomial output. The insets show the intensities of the input and output beams while the panels show the beam profiles.
	}
	\label{fig:DickeyMap}
\end{figure}
In Fig.~\ref{fig:DickeyMap}, a truncated Gaussian input is mapped to an annular polynomial output defined by the equation $X^2 - X^4$. Note the following: the function $\alpha(x)$ maps points from the input plane to points of the output plane; the points near the centre of the input plane where the beam is most intense indicated by the red solid and magenta dotted lines are mapped to points further apart on the output plane than a point further out on the input plane; the function $\alpha(x)$ has a maximum value $\omega_{\text{opt}}$, corresponding to the waist of the output beam, and, the scaling constant $A$ as given in Eqn.~\ref{eqn: Aformula} relates the shaded area of the output profile to the shaded area of the input profile ensuring that the method is lossless since energy is conserved.
\subsection{Properties of the $\alpha(x)$ function}
The mapping function $\alpha(x)$ has three properties of interest. Firstly, Eqn.~\ref{eqn: Cartesian Alpha} breaks down for $|Q(\alpha(x))|^2 = 0$. The formulation of $\alpha(x)$ cannot map an input beam to a region of the output beam containing no light as this would require a phase singularity. Secondly, and as a consequence of the first property, the value of $\alpha(x)$ monotonically approaches an asymptotic maximum value corresponding to the value of the waist of the output beam $\omega_{\text{opt}}$ as seen in Fig.~\ref{fig:DickeyMap}. Lastly the choice of coordinate system has a significant impact on the structure and functioning of $\alpha(x)$.\\
\\ 
Consider a circular Gaussian beam in Cartesian separated coordinates $|\exp\left(-\frac{x^2}{2}\right)\exp\left(-\frac{y^2}{2}\right)|^2$, and in Polar coordinates $|\exp\left(-\frac{r^2}{2}\right)|^2$. When integrating over all space the results of these two expressions are the same. However, if we limit the working to only a one dimensional integral, along $x$ in the Cartesian case and $r$ in the Polar case, the result is $\sqrt{\pi}$ and $\frac{1}{2}$ respectively. This difference in scaling effects $A$, the mapping function $\alpha$ and then by extension the phase element. Functionally, by reducing the problem to 1-dimension for simplified solving a 'propagation of coordinate system' is inadvertently introduced throughout the equations. If the Cartesian expression is chosen, the phase element will attempt to produce an output profile with Cartesian symmetry, while if the Polar expression is used, the output profile will have a radial symmetry.  

\subsection{Deriving the second phase element}
Once the first phase element has been designed there are two viable approaches to designing the second phase element. For problems where the $\beta$ parameter can be engineered to be large the second phase element can be well approximated as the complex conjugate of the first shaping element. Alternatively, if the $\beta$ parameter is small the second element can be better designed by using numerical simulation to determine the phase of the shaped beam at the relevant plane.

\section{Experimental Implementation}
\begin{figure}[H]
	\centering
	\includegraphics[width=.5\textwidth]{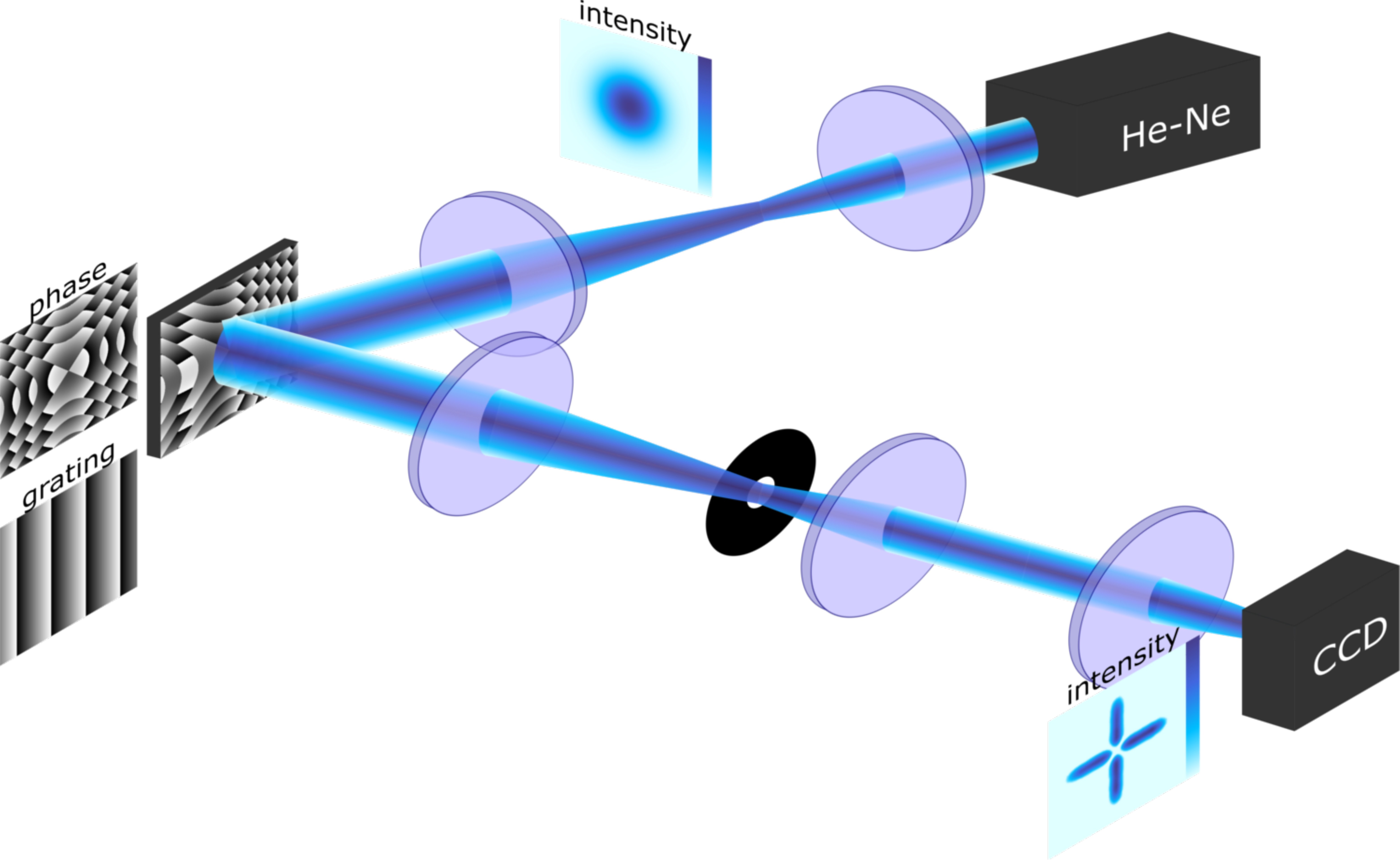}
	\caption{Design of the experimental setup used to implement the method. Here, the second beam shaping element described in the introduction would be placed at the plane of the black aperture.
	}
	\label{fig:DickeySetup}
\end{figure}

To demonstrate the method experimentally we chose to convert a Gaussian beam into a number of different spatial profiles. This choice was made because of the ubiquity of laser sources which output Gaussian modes. Figure~\ref{fig:DickeySetup} shows a diagram of the experimental setup used. The first phase element, as described in the introduction, was modelled using a Holoeye Pluto 1 Spatial Light Modulator (SLM) illuminated by a well expanded and collimated beam from a HeNe laser source. The SLM displayed a hologram which created a beam with a Gaussian amplitude profile and a phase calculated by Eqn.~\ref{eqn: dickey phase eqn}. The beam from the SLM passed through a lens. At the focal plane of this lens the desired output beam $|Q(\rchi)|^2$ formed. This mode was filtered and imaged to a CCD.

In this experiment we were primarily concerned with measuring the amplitude of the beam generated by the method, as such we did not implement a second phase element. The motivation for this is two fold. Firstly, the derivation of the first phase element is considerably more complex than the derivation of the second and so experimentally validating the design of the first element held priority. Secondly, the inclusion of the second element in the experimental setup would not have impacted the intensity of the beam in a measurable way, instead changing only the phase.

\section{Experimental Results}
\begin{figure}[H]
	\centering
	\includegraphics[width=0.45\textwidth]{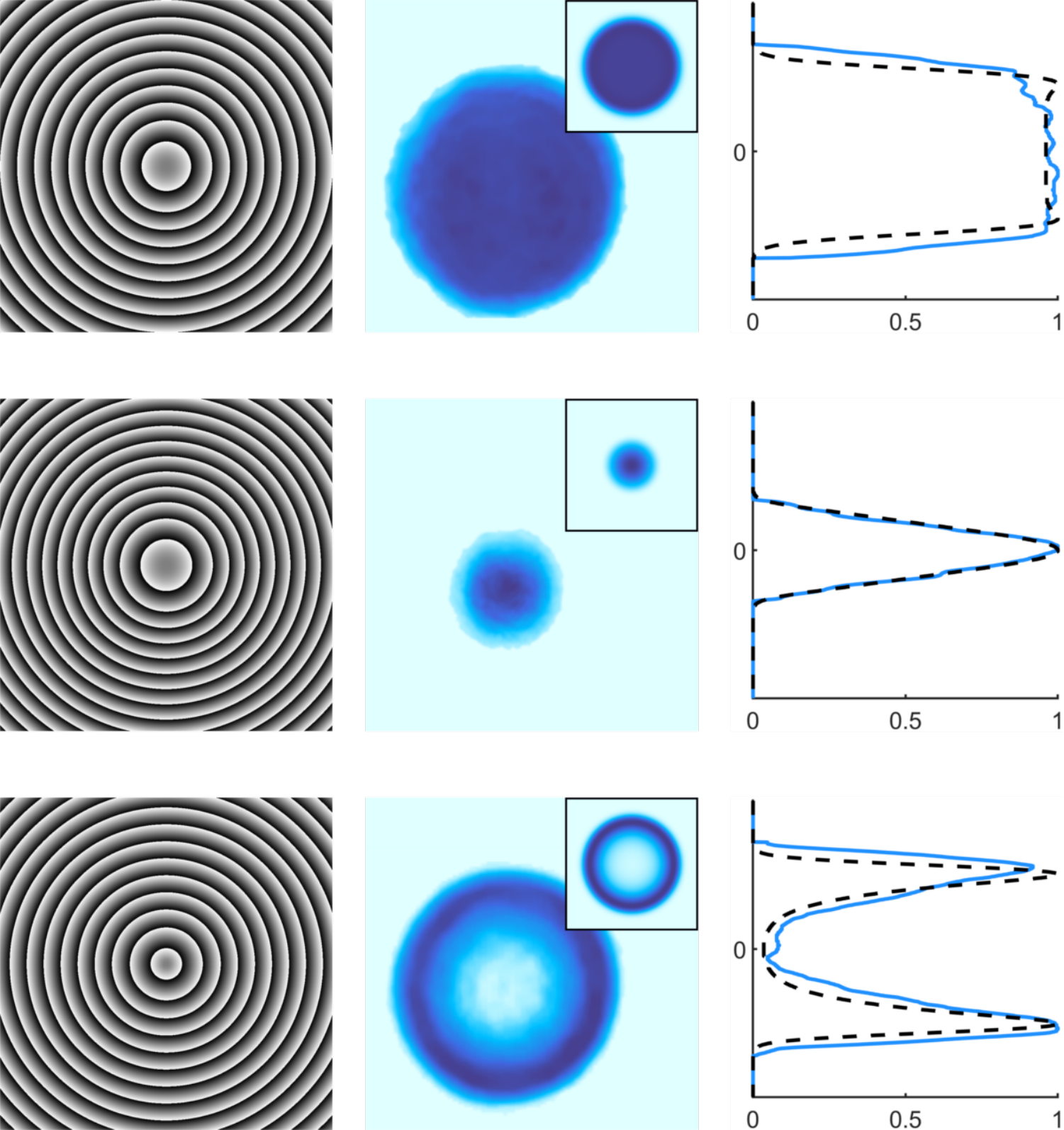}
	\caption{Measured modes and profiles for beams with Radial symmetry. Each row of the figure shows from left to right, the phase of the first shaping element, the experimentally measured beam, and a comparison of the profiles of the simulated and measured beams as a black dashed line and blue solid line respectively. The insets in each experimental result show the simulated beam.}
	\label{fig:RadialResults}
\end{figure}

\begin{figure}[t]
	\centering
	\includegraphics[width=0.45\textwidth]{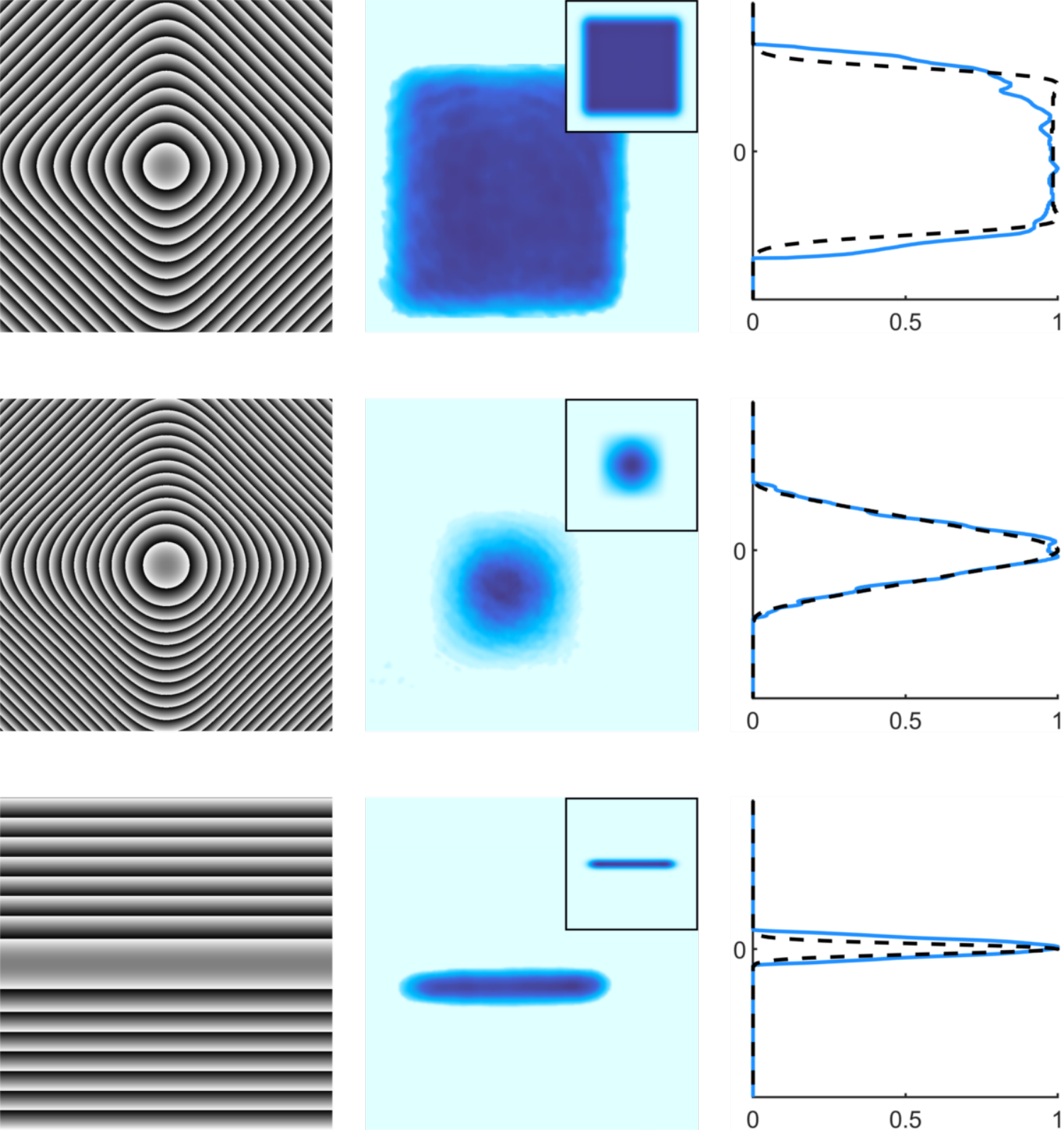}
	\caption{Measured modes and profiles for beams with Cartesian symmetry. Each row of the figure shows from left to right the phase of the first shaping element, the experimentally measured beam, and a comparison of the profiles of the simulated and measured beams as a black dashed line and blue solid line respectively. The insets in each experimental result show the simulated beam.}
	\label{fig:CartesiaResults}
\end{figure}

The experimental results shown in Fig.~\ref{fig:RadialResults} and Fig.~\ref{fig:CartesiaResults} were created using the experimental setup represented by Fig.~\ref{fig:DickeySetup}. The beams were generated by encoding a Gaussian beam of waist $\omega_G$ with the appropriate phase to form an output profile of waist $\omega_{\text{opt}}$ at the focal plane of a lens with focal length $0.5$ m. The output profile was then captured by a camera. The waist parameters and cameras used are summarised in Tab.~\ref{tab:ExpBeamParam}. 
\begin{table}[h]
	\centering
	\caption{\bf The cameras, Gaussian and output profile waists used for the experiment}
	\begin{tabular}{cccc}
		\hline
		\textbf{Profile} & \textbf{$\omega_G$ (m)}& \textbf{$\omega_{\text{opt}}$ (m)}& \textbf{Camera}\\
		\hline
		\text{Radial flat-top} &\multirow{3}{*}{0.0011}&\multirow{3}{*}{0.0008}&\multirow{4}{*}{Logitech C270}\\
		\text{Radial Annulus}&&&\\ 
		\text{Cartesian 2D flat-top }&&&\\
		\cline{1-3}
		\text{Cartesian 1D Gaussian}& \text{0.0011} &\text{0.0007}&\\
		\hline
		\text{Radial Linear}&\multirow{2}{*}{0.00087}&\multirow{2}{*}{0.0011}&\multirow{2}{*}{Thorlabs 1240C}\\
		\text{Cartesian Linear}&&\\
		\hline		
	\end{tabular}
	\label{tab:ExpBeamParam}
\end{table}
In the case of the Logitech C270 camera, all optics of the camera were removed to leave the sensor exposed.\\
\\
Figure~\ref{fig:RadialResults} shows the measured beams having radial symmetry. The phase of the first shaping element, the experimentally measured beam (as well as an inset of the simulated beam) and a comparison of the profiles of the measured and simulated beams are shown. The beams shown in Fig.~\ref{fig:RadialResults} are a flat-top, a beam with a linear intensity profile, and an annular beam. Figure~\ref{fig:CartesiaResults} displays the same but for beams of Cartesian symmetry. The beams shown in Fig.~\ref{fig:CartesiaResults} are a flat-top, a beam with a linear intensity profile, and a 1-dimensional Gaussian profile where the beam has a focused Gaussian intensity across one axis of its profile and a uniform profile in the other axis.\\
\\
To quantify the general fidelity of all the generated modes the native \textit{Corr2} function of \textit{Matlab} was used. The correlation values between the simulated and experimentally generated modes are presented in Tab.~\ref{tab:BeamCorrelation}.
\begin{table}[h!]
	\centering
	\caption{\bf The correlations between the simulated and measured beams}
	\begin{tabular}{cc}
		\hline
		\textbf{Profile} & \textbf{Correlation} \\
		\hline
		\text{Radial flat-top} & \text{0.951} \\
		\hline
		\text{Radial Annulus} & \text{0.927}\\
		\hline
		\text{Radial Linear} & \text{0.991}\\
		\hline
		\text{Cartesian 1D Gaussian} & \text{0.909} \\
		\hline
		\text{Cartesian 2D flat-top} & \text{0.956} \\
		\hline
		\text{Cartesian 2D Linear} & \text{0.995} \\
		\hline			
	\end{tabular}
	\label{tab:BeamCorrelation}
\end{table}
\begin{figure}[H]
	\centering
	\includegraphics[width=.40\textwidth]{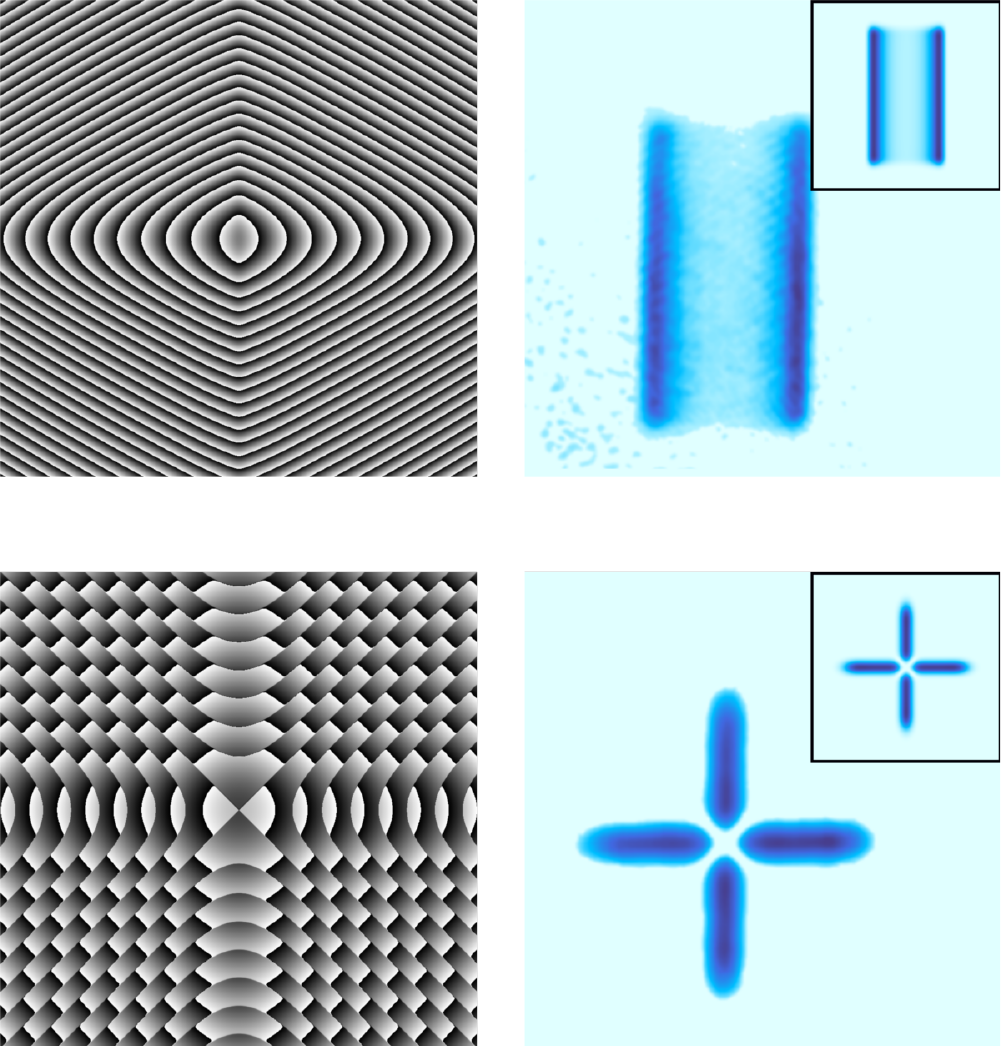}
	\caption{The 'double bar' and 'cross hair' profiles. These profiles are the result of superpositions of modes at right angles to each other in a Cartesian symmetry.}
	\label{fig:SuperPosResults}
\end{figure}
The results presented here demonstrate how the method can be used to generate modes with unique spatial profiles while maintaining high fidelity. An interesting extension to the method we present is how it may be applied to superpositions of modes.\\
\\
In particular the separable nature of the Cartesian coordinate system allows two separate profiles to be encoded perpendicularly to one another. Figure~\ref{fig:SuperPosResults} shows the 'double bar' structure that emerged as a combination of a flat-top profile and a linear valley profile at right angles. Figure~\ref{fig:SuperPosResults} also shows the combination of two 1D Gaussian profiles at right angles which gives rise to a 'cross hair' type mode.

\section{Conclusion}
In this work we have demonstrated a general approach to the reshaping of structured light. The approach addresses the concept of lossless two step beam shaping, highlighting how two or more phase only elements can perfectly reshape light, something often not achieved when using existing well established beam shaping techniques. Although based on prior work we have expanded the method beyond its well-known capability of converting Gaussian beams into flat-top beams. Our general formulation allows for the conversion of almost any input beam into a wide range of spatial profiles having both a specific phase and amplitude. This generality is supported by the experimental results we present. Our experimental results also demonstrate the feasibility of implementing the method in the laboratory environment. Critically, the work we present is not limited to the spatial shaping of \textit{scalar} light, but can form a foundation to be used in conjunction with temporal and vectorial shaping techniques to provide complete control over all the characteristics of light. Thus, the work presented here represents a powerful tool for the lossless reshaping of light to fit a myriad of applications.

\section{Acknowledgements}
The authors would like to thank the Council of Scientific and Industrial Research with the Department of Science for providing funding through the Interbursary Incentive Funding programme (CSIR-DST IBS).\\

\section{Disclosures}
\noindent\textbf{Disclosures.} The authors declare no conflict of interest in the production or publication of this work.

\bigskip
\noindent


\end{document}